# Chemical Stability of Laponite in Aqueous Media

Shweta Jatav and Yogesh M Joshi [a]

Department of Chemical Engineering, Indian Institute of Technology Kanpur 208016, India

[a] Corresponding author, E-Mail: joshi@iitk.ac.in, Phone: 91-512-2597993

**Abstract**

In this work stability of Laponite against dissolution in its aqueous dispersions is investigated as a function of initial pH of water before mixing Laponite, and concentration of Laponite. Dissolution of Laponite is quantified in terms of concentration of leached magnesium in the dispersions. Interestingly the solvent pH is observed to play no role in dissolution of Laponite in dispersion over the explored range of 3 to 10. Furthermore, contrary to the usual belief that Laponite dissolves when pH of aqueous dispersion decreases below 9, dissolution of the same is observed even though dispersion pH is above 10 for low concentrations of Laponite (1 and 1.7 mass%). On the other hand, for dispersions having high concentration of Laponite (2.8 mass%) and pH in the similar range (>10) no dissolution is observed. Measurement of ionic conductivity of dispersion shows that concentration of sodium ions in dispersion increases with concentration of Laponite, which appears to have a role in preventing the dissolution of Laponite.

## 1. Introduction

Laponite, a synthetic clay mineral, is known to have widespread applications as a rheology modifier and as a reinforcement in variety of industries such as mining, petroleum, home and personal care, pharmaceutical, agrochemical, paint polymer, etc. Primary particle of Laponite possesses anisotropic nanometric shape that has dissimilar charge distribution. Consequently its dispersion in water shows a rich variety of phase behaviours (Mourchid et al., 1995; Mongondry et al., 2005; Jabbari-Farouji et al., 2008; Ruzicka and Zaccarelli, 2011; Shahin et al., 2011; Shahin and Joshi, 2012; Sun et al., 2012; Tudisca et al., 2012). In addition, it attracts applications as an active agent in many water based formulations (Negrete et al., 2004; Sun et al., 2009; Ghadiri et al., 2013). The dependence of physical properties of Laponite dispersion on time, particularly the observed increase in modulus and relaxation time, is reminiscent of physical aging in molecular and spin glasses (Schosseler et al., 2006; Shahin and Joshi, 2011; Morariu and Bercea, 2012; Dhavale et al., 2013). Owing to this, aqueous dispersion of Laponite is also investigated as a model soft glassy material (Bonn et al., 2002; Bandyopadhyay et al., 2004). Over the past two decades Laponite in aqueous dispersion as well as in other multi-component systems has attracted enormous attention from the academia as well as from the industry. However, Laponite is reported to have a major shortcoming related to its chemical stability. According to Thompson and Butterworth (1992), Laponite particles undergo dissolution in the aqueous media having pH less than 9. Interestingly despite vast literature available on this clay mineral, very few papers study the chemical stability of Laponite dispersion, which is a subject of this work. It is observed that Laponite particles in dispersion are prone to dissolution even at high pH and chemical stability of the same strongly depends on the concentration of Laponite.

Laponite has a chemical formula given by: $Na_{0.7}Si_8Mg_{5.5}Li_{0.3}O_{20}(OH)_4$. Laponite particles are disk shaped with thickness 1 nm and diameter $25\pm2$ nm (Kroon et al., 1998). In a single layer of Laponite two tetrahedral silica sheets sandwich one octahedral magnesia sheet. In the middle octahedral sheet few magnesium atoms are substituted by lithium atoms (isomorphic substitution) creating deficiency of positive charge within the sheet. Consequently, in a dry state, the faces of Laponite particles, that are electron rich, share the electrons with sodium atoms that reside in the interlayer space. Upon dispersing in the aqueous media the $Na^+$ ions dissociate rendering a permanent negative charge to the faces of Laponite particles. The edge of Laponite particle predominantly contains MgOH groups from the octahedral magnesia sheets. The point of zero charge (PZC), for oxides and hydroxides of magnesium is above pH of 10 (Kosmulski, 2001). Martin et al. (2002), mentioned that according to the manufacturer (Laponite Technical Bulletin, 1990) the edge of Laponite particle, which contains predominantly MgOH, is positive below pH of 11 indicating pH of 11 to be point of zero charge for the edges of Laponite particles. Depending upon the pH of the medium, either $H^+$ or OH ions dissociate from the edges rendering the same negative or positive charge respectively. Dissociation of $H^+$ ions from the edge occurs only above the pH associated with PZC to acquire the negative charge. Consequently pH of the dispersion decreases. Below the pH associated with PZC, the edge of Laponite particle releases $OH^-$ ions, which causes increase in pH of the dispersion. Therefore, depending upon whether $H^+$ or OH ions dissociate from the edge, the pH of dispersion respectively decreases or increases (Tawari et al., 2001), and the resultant value of the pH of Laponite dispersion has strong influence on the stability of the same.

According to available literature, the first study on chemical stability of hectorite in aqueous media is due to Tiller (1968). He estimated leaching of magnesium ions from naturally occurring purified hectorite using atomic absorption spectroscopy. He observed that concentration of leached magnesium increases with decrease in pH but remains practically independent of concentration over the explored range of 0.1 to 0.4 mass%. The first investigation on chemical stability of Laponite, which is a synthetic hectorite, in aqueous dispersion was performed by Thompson and Butterworth (1992). They systematically studied the effect of the pH of the medium on stability of Laponite dispersion in the low concentration regime (below 2 mass%). They observed detectable dissolution of Laponite below the pH of 9. They proposed that such dissolution follows a chemical reaction given by:

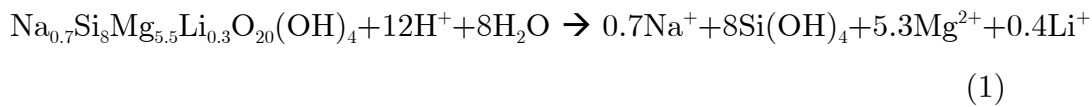

$$Na_{0.7}Si_8Mg_{5.5}Li_{0.3}O_{20}(OH)_4 + 12H^+ + 8H_2O \rightarrow 0.7Na^+ + 8Si(OH)_4 + 5.3Mg^{2+} + 0.4Li^+$$

$$(1)$$

wherein excess of $H^+$ ions causes leaching of magnesium ions from the particles. There is a slight difference between formula for Laponite mentioned by Thompson and Butterworth (1992) and that adapted in this work. Thompson and Butterworth suggested formula to be $Na_{0.8}Si_8Mg_{5.4}Li_{0.4}O_{20}(OH)_4$, while this work subscribes to the formula suggested by the manufacturer. The above reaction therefore has been stoichiometrically corrected to suit Laponite used in this study.

Mourchid and Levitz (1998) studied long term gelation of aqueous dispersion of Laponite having low concentration of Laponite (1 and 1.5 mass%). They observed that samples preserved under inert conditions (nitrogen atmosphere in this case) do not undergo dissolution (do not show any traces of $Mg^{+2}$ ions). On the other hand, the samples that are merely sealed indeed show presence of $Mg^{+2}$ ions, which increases as a function of time. They report that all the samples wherein concentration of $Mg^{+2}$ ions is observed to be greater than 0.5 mM form viscoelastic gels. They claim that dissolution of atmospheric $CO_2$ is sufficient to create acidic environment which according to reaction (1) leads to leaching of $Mg^{+2}$ ions. Interestingly Mourchid and Levitz (1998) prepare the dispersions of Laponite in water having pH 10. However they do not report the pH of dispersion when presence of $Mg^{+2}$ ions was observed. Apparently studies by Thompson and Butterworth (1992) and Mourchid and Levitz (1998) are the only two reports available in the literature on the chemical stability of Laponite dispersion. Both the reports suggest possibility of leaching of $Mg^{+2}$ ions when the pH of dispersion is below 9.

## 2.      Materials and Experimental Procedure

Laponite XLG® used in this study is obtained from Southern Clay Products Inc. Laponite is dried for 4 h at 120ºC to remove the moisture and is subsequently mixed with Millipore water (Resistivity= 18.2 MΩ.cm) having pH in the range 3 to 10. For maintaining the acidic pH two reagents have been used, namely: citric acid buffers as well as HCL. Citric acid buffers are prepared by incorporating 0.1 M citric acid and 0.1 M tri-sodium citrate solutions in water. In order to maintain the basic pH, NaOH has been employed. In some samples, NaCl has also been incorporated. Once the predetermined initial pH of water and salt concentration are obtained the dried Laponite powder is added to the same. Mixing is carried out using an Ultra Turrex drive for a period of 45 min. The dispersions are then stored in sealed polypropylene bottles without any nitrogen purging. After the sample preparation bottles are filled in such fashion that they have around 200 ml of open space filled with air above the sample. However, the bottles are opened in order to take out the sample for pH and $Mg^{+2}$ ion concentration measurements at predefined interval of days (duration between two consecutive samples vary between 1 to 5 days). The list of samples studied in this work with respect to concentrations of Laponite and initial pH is reported in Table 1. Throughout in this paper pH of water before mixing Laponite is represented as $pH_i$. On the other hand, pH of dispersion (after addition of Laponite) is termed as $pH_f$, which is observed to depend on time.

**Table 1.** List of samples studied for chemical stability against $Mg^{+2}$ ions leaching

| Laponite Conc. (mass %) | Acidic $pH_i$ (No Salt) | Neutral and basic $pH_i$ (No Salt) | Concentration of NaCl (mM) with $pH_i$ =7 |
|---|---|---|---|
| 1 | -- | $pH_i$ =7, 8, 9 and 10. Basic $pH_i$ maintained by NaOH | 11.1 |
| 1.7 | -- | | 7.2 |
| 2.8 | $pH_i$ =3, 4 and 6 maintained by Citric acid buffer as well as HCL | | -- |

In order to detect concentration of $Mg^{+2}$ ions in Laponite dispersion, complexometric titration is performed on dispersion samples at regular interval after preparation of the same. Complexometric titration is performed with EDTA using eriochrome black-T as an indicator. If dispersion contains $Mg^{+2}$ ions, is turns red or purple upon addition of eriochrome black-T indicator. The method is so sensitive that it can detect concentration of $Mg^{+2}$ ions as small as $10^{-3}$ mM (Vogel, 1978). The details of complexometric titration procedure can be found elsewhere (Vogel, 1978). It is important to note that Laponite dispersion forms a high viscosity/elasticity gel, and therefore, in principle, titration of the same in this form is difficult. However, Laponite gel is thixotropic, therefore its viscosity/elasticity can be reduced significantly by simply shearing it vigorously. Therefore, it has been made sure that the gel viscosity has reduced significantly by shearing it to the extent possible so that it is in liquid state at the time of titration. The changes in ionic conductivity and $pH_f$ also measured as a function of time using Eutech Cyberscan CON 6000 pH and conductivity meter with 4 cell conductivity electrode (range 0-500 mS and temperature range 0-70°C) and an open pore double reference junction Ag/AgCl pH electrode (range pH 0-14 and temperature 0-80°C). All the experiments are performed at 25°C.

# 3.    Results and Discussion

Upon incorporating Laponite in water the nature of resultant dispersion is strongly influenced by $pH_i$ and reagents used to maintain the $pH_i$. Particularly, 2.8 mass% dispersions have been prepared in water with $pH_i$ between 3 and 10. As mentioned before, in order to maintain acidic $pH_i$, two reagents namely citric acid buffer and HCl are used. It is observed that when $pH_i$ is maintained (in the range 3 to 10) by using either HCl or NaOH, dispersions of 2.8 mass% Laponite are transparent and eventually acquire a soft solid like consistency. On the other hand, when citric acid buffer is used to maintain $pH_i$ of 3 and 4, resultant 2.8 mass% dispersions show sedimentation wherein Laponite settles down within a day. However, for $pH_i$ =6 (obtained by citric acid buffer) 2.8 mass% dispersion does not undergo sedimentation, but becomes hazy and remains in the liquid state throughout the observation period of 30 days. In addition, 1 and 1.7 mass% dispersions are also prepared in water having $pH_i$ between 7 to 10. It is observed that 1 mass% dispersion remains in liquid state without noticeable change in viscosity over a period of 30 days. Dispersions having 1.7 mass% concentration also remain in liquid state but become progressively more viscous over the observation period of 30 days.

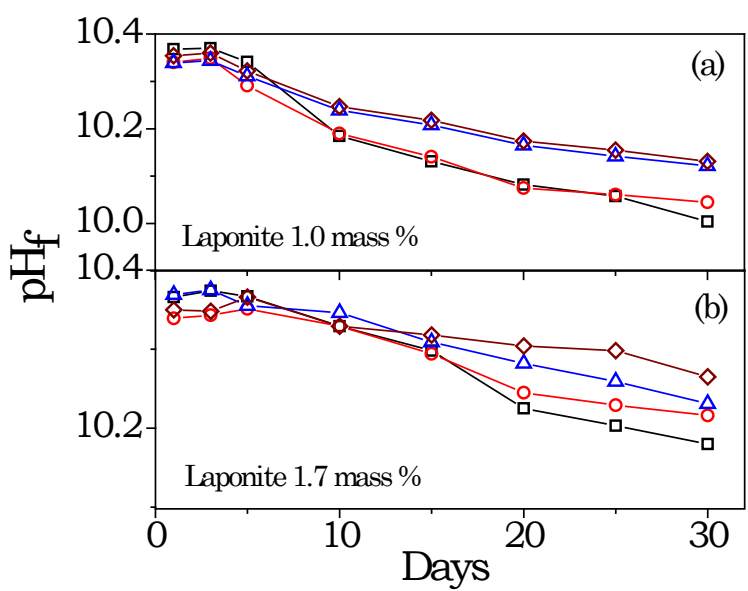

**Figure 1** Evolution of $pH_f$ of (a) 1 mass% and (b) 1.7 mass% Laponite dispersion plotted as a function of number of days after preparing the same in water having $pH_i$ =7 (squares), 8 (circles), 9 (up triangles) and 10 (diamonds). The lines serve as a guide to the eye.

Incorporation of Laponite in water, depending upon $pH_i$, leads to dissociation $OH^-$ ions from its edge. Such dissociation in turn causes increase in $pH_f$ of the dispersion. In figures 1 and 2 evolution of dispersion $pH_f$ is plotted as a function of number of days since preparation of the same for different concentrations of Laponite. For 1 and 1.7 mass% Laponite dispersions with $pH_i$ in the range 7 to 10, time dependent evolution of $pH_f$ is shown in figure 1a and 1b respectively. It can be seen that soon after mixing Laponite, $pH_f$ increases to values above 10. As time passes, $pH_f$ decreases weakly as a function of time, so that over a period of 30 days $pH_f$ still remains above 10. Furthermore, lesser decrease in $pH_f$ is observed in dispersions prepared in water with greater $pH_i$.

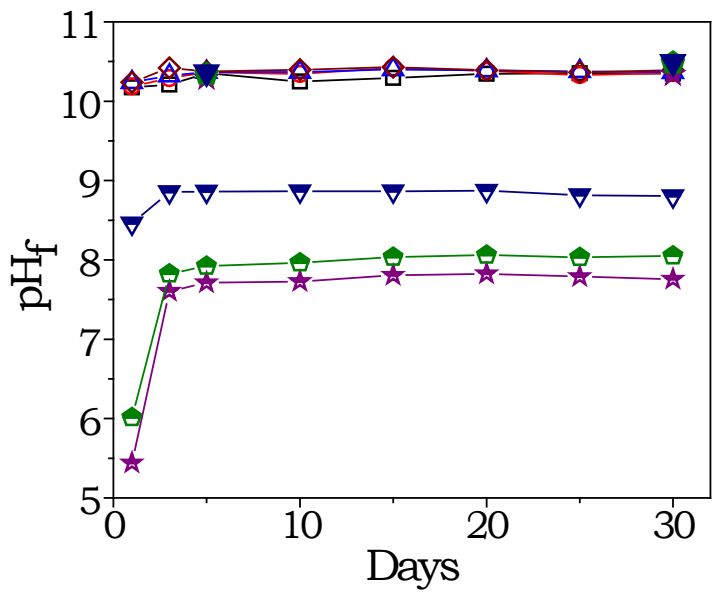

**Figure 2** Evolution of $pH_f$ of 2.8 wt % Laponite dispersion plotted as a function of number of days after preparing the same in citric acid buffer solution (half filled

symbols) having $pH_i = 3$ (stars), 4 (pentagons), 6 (down triangles); in HCl solution (filled symbols) having $pH_i = 3$ (stars), 4 (pentagons), 6 (down triangles); and in NaOH solution (open symbols) having $pH_i = 7$ (squares), 8 (circles), 9 (up triangles) and 10 (diamonds)]. The lines serve as a guide to the eye.

Variation in $pH_f$ for 2.8 mass% Laponite dispersion for $pH_i$ in the range 3 to 10 is shown in figure 2. For 2.8 mass% Laponite dispersion with acidic $pH_i$ (3 to 6) maintained by citric acid buffer, $pH_f$ increases within first 5 days and then remains constant. The value of constant $pH_f$ always remains below 9. On the other hand, if the acidic $pH_i$ (3 to 6) is maintained by HCL, addition of 2.8 mass% Laponite causes increase in $pH_f$, which also stabilizes within first 5 days and remains nearly constant at value of 10.4 irrespective of value of $pH_i$. For 2.8 mass% dispersions with $pH_i$ in the range 7 to 10 (basic $pH_i$ maintained by NaOH) also the increase in their $pH_f$ values are observed. However regardless of $pH_i$, $pH_f$ is observed to reach a constant level of around 10.4 after 5 days. Interestingly for 2.8 mass% Laponite dispersion having $pH_i$ in the range 3 to 10 maintained by either HCL or NaOH, $pH_f$ after 5 days is observed to be very similar and around 10.4. It should be noted that below a point of zero charge (PZC), which is around pH of 11 for the edge of Laponite particle (Martin et al., 2002), edges acquire positive charge by dissociating $OH^-$ ions. Such dissociation of $OH^-$ ions raises the pH of the aqueous media in which Laponite particles are dispersed in. Since dissociation of $OH^-$ ions from the edges of Laponite particle depends upon the pH of the surrounding aqueous medium and not the initial pH, at the pH of 10.4 the dissociation from the edge stops and pH does not increase any further.

In figure 3, comparison of the results of figure 1 and 2 are shown, wherein $pH_f$ on day 5 plotted as a function of $pH_i$. It can be seen that dispersions prepared in citric acid buffer show qualitatively different behaviour than the rest of the dispersions. This different behaviour can be understood as follows. In aqueous solution, incorporation of incremental amount of acid or base (or a compound that tends to release $OH^-$ or $H^+$ ions) tends in to change its $pH_f$ gradually. In a buffer, on the other hand, a mixture of weak acid (in this case citric acid) and its conjugate base (in the present case tri sodium citrate) is used, which causes only a small change in $pH_f$ even when strong acid or base is incorporated in the same. Buffer solutions tend to resist change in $pH_f$ owing to equilibrium between acid and its conjugate base (Bettelheim et al., 2009). This aspect is very evident from figure 2. Consequently $pH_f$ of Laponite dispersion prepared in citric buffer always remains below 9.

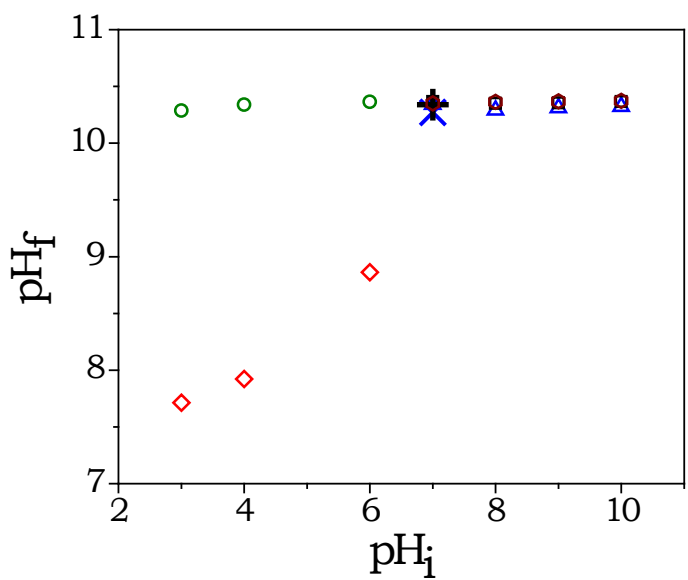

**Figure 3** $pH_f$ of dispersion on day 5 is plotted as a function of $pH_i$ at different concentrations: 1 mass% (up triangles: no salt, multiplication sign (×): 11.1 mM NaCl), 1.7 mass% (squares: no salt, plus sign (+): 7.2mM NaCl), 2.8 mass% with $pH_i$ maintained by HCl or NaOH (circles) and citric acid buffers (diamonds).

The change in $pH_i$ upon incorporation of Laponite in water has an important consequence as the resultant $pH_f$ of Laponite dispersion, as reported in the literature, is proposed to affect the chemical stability of the Laponite particles in the dispersion. As discussed before, in the presence of $H^+$ ions, Laponite particles undergo reaction (1), which results in leaching of magnesium ($Mg^{+2}$) and lithium ($Li^+$) ions. In this work, estimation of the concentration of $Mg^{+2}$ ions is carried out by complexometric titration of Laponite dispersions at regular intervals. The corresponding measured concentration of $Mg^{+2}$ ions for three concentrations of Laponite as a function of number of days since preparation of the same is plotted in figure 4. Laponite dispersions having 2.8 mass% concentration with acidic $pH_i$ obtained by citric acid buffer (and $pH_f < 9$ for all samples) shows significant amount of leaching of $Mg^{+2}$ ions, whose magnitude also increases with time as shown in figure 4(a). For concentrations 1 and 1.7 mass% and $pH_i = 7$ to 10 (and $pH_f > 10$ for all samples) also shows measurable presence of $Mg^{+2}$ ions at very early age, which increases with time. Interestingly magnitude of $Mg^{+2}$ ions present in 2.8 mass% dispersion with $pH_i$ maintained by citric acid buffer is observed to be significantly greater in magnitude than lower concentration dispersions whose $pH_i$ is maintained by NaOH. However, most surprisingly, for 2.8 mass% Laponite dispersion with $pH_i = 3$ to 10 maintained by adding HCl or NaOH, no traces of $Mg^{+2}$ ions are observed over the explored duration of 30 days.

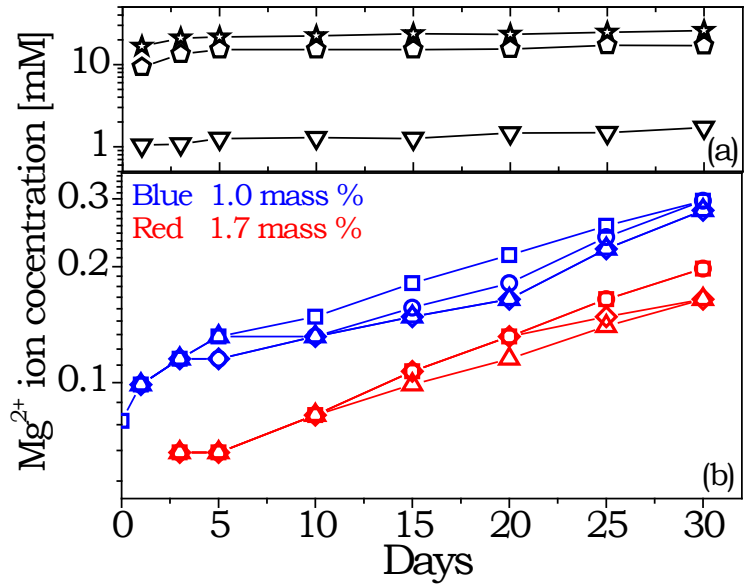

**Figure 4** Concentration of $Mg^{2+}$ ions in Laponite dispersion plotted as a function of days elapsed since preparation of dispersion for (a) 2.8 mass% dispersion with $pH_i$ maintained by citric acid buffers, and (b) 1.0 and 1.7 mass% dispersion. The symbols

represent, star: $pH_i=3$, pentagon: $pH_i=4$, down triangle: $pH_i=6$, square: $pH_i=7$, circle: $pH_i=8$, up triangle: $pH_i=9$ and diamond: $pH_i=10$. The lines serve as a guide to the eye.

As mentioned in the introduction, the very first report on chemical stability of Laponite by Thompson and Butterworth (Thompson and Butterworth, 1992) claims that particles of Laponite undergo dissolution leading to leaching of $Mg^{+2}$ ions when $pH_f$ decreases below 9. However the present study clearly demonstrates that even though the $pH_f$ is above 10, leaching of $Mg^{+2}$ ions does take place in the dispersions having low Laponite concentration. Surprisingly the dispersions with 2.8 mass% concentration, although has $pH_f$ in the similar range as that of the low concentration dispersions, do not show any measurable presence of $Mg^{+2}$ ions. Even though the overall $pH_f$ is basic, as suggested by Mourchid and Levitz (1998), dissolution of atmospheric $CO_2$ in water produces carbonic acid and hence $H^+$ ions locally according to (Greenwood and Earnshaw, 1997):

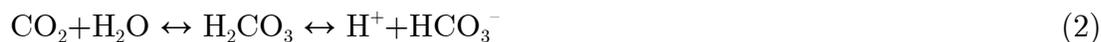

$$CO_2 + H_2O \leftrightarrow H_2CO_3 \leftrightarrow H^+ + HCO_3^- \tag{2}$$

Therefore continuous enhancement of $Mg^{+2}$ ions reported in figure 4 can be attributed to slow dissolution of $CO_2$ from the atmosphere that follows reactions (1) and (2).

The reason of this significant leaching of $Mg^{+2}$ ions in 2.8 mass% Laponite dispersion ( $pH_i$ controlled by citric acid buffer) is already low $pH_f$ of dispersion compared to all the other dispersion samples aided by dissolution of atmospheric $CO_2$. Dissolution of Laponite in low concentration dispersion with $pH_i$ in the range 7 to 10 and $pH_f \approx 10.4$ can be ascribed to dissolution of atmospheric $CO_2$, which leads to generation of $H^+$ ions locally. As soon as $H^+$ ions are generated, there is competition between Laponite particles and $OH^-$ ions (associated with the basic environment). If $H^+$ ions react with Laponite it will lead to dissolution. On the other hand, if $H^+$ ions react with $OH^-$ ions, it will cause decrease in pH. (In this work, as mentioned in Table 1, 1 and 1.7 mass% dispersions in water having acidic $pH_i$ are not studied. Since even 1 and 1.7 mass% dispersions prepared in basic $pH_i$ demonstrate leaching of $Mg^{+2}$ ions, it is obvious that 1 and 1.7 mass% dispersions in water having acidic $pH_i$ will also show dissolution of Laponite particles.) The stability of 2.8 mass% Laponite dispersion where $pH_i$ is controlled by HCL or NaOH (with $pH_i = 3$ to 10 and $pH_f \approx 10.4$) observed in the third case is however puzzling.

It is known that $CO_2$ has significant affinity towards NaOH, and dissolved $CO_2$ reacts with NaOH to first form sodium bicarbonate ($CO_2+NaOH \rightarrow NaHCO_3$) and finally sodium carbonate ($NaHCO_3+NaOH \rightarrow Na_2CO_3+H_2O$). Therefore greater concentration of NaOH in solution/dispersion is expected to inhibit the formation of carbonic acid and $H^+$ ions, and in turn prevent leaching of the $Mg^{+2}$ ions. However, for 1, 1.7 and 2.8 mass% dispersions (with $pH_i$ maintained either by HCl or NaOH), the $pH_f$ is always around 10.4 over the explored period as shown in figures 1 to 3. Consequently concentration of solvated NaOH is also the same in all these cases. This suggests that effects originating from higher concentration of Laponite are able to prevent reaction of $CO_2$ and $H_2O$ that produces $H^+$ ions.

In order to investigate effect of Laponite concentration further, the ionic conductivity measurements of 1 and 1.7 mass% Laponite dispersions ($pH_i = 7$ to 10 maintained by NaOH) and 2.8 mass% Laponite dispersion ($pH_i = 3$ to 10 maintained by HCl or NaOH) are also carried out. In figure 5 ionic conductivity ($\sigma$) as a function of $pH_i$ is plotted (filled symbols) for different concentrations of Laponite. The reported conductivity is measured on day 5 after preparing the dispersion. It can be seen that ionic conductivity of solvent as a function of $pH_i$ shows a minimum at 7. The corresponding ionic conductivity of Laponite dispersion, however, seems to be independent of $pH_i$ over the explored range but shows greater increase for higher concentration of Laponite. The reason behind enhancement in conductivity of water having certain $pH_i$ after addition of Laponite is primarily due to dissociation of $Na^+$ counterions from the faces of the Laponite particles. Furthermore as shown in figure 3, incorporation of Laponite in water having different $pH_i$ causes increase in the $pH_f$ of the dispersion suggesting dissociation of $OH^-$ counterions. Such dissociated $OH^-$ counterions from the edges are also responsible for increase in ionic conductivity shown in figure 5.

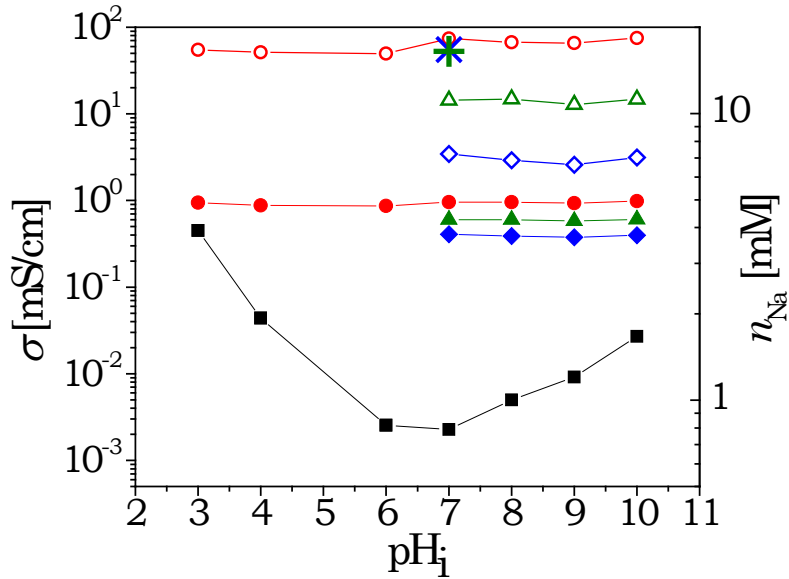

**Figure 5** Ionic Conductivity, $\sigma$ (filled symbol) and sodium ion concentration, $n_{Na}$ (open symbols) are plotted as a function of $pH_i$ (maintained either by HCl or NaOH) after 5 days of preparation of dispersion for 1 (diamond: no salt, multiplication sign ($\times$): 11.1 mM NaCl), 1.7 (up triangle: no salt, plus sign (+): 7.2 mM NaCl), 2.8 (circle) mass% Laponite dispersion and solvent conductivity (squares). The lines serve as a guide to the eye.

The ionic conductivity ($\sigma$) of a solution/dispersion is related to concentration of $i$th type of ion ($n_i$) present in the same by (Benenson et al., 2002):

$$\sigma = e \sum_i \mu_i n_i, \tag{3}$$

where $\mu_i$ is mobility of $i$th type of ion and $e$ is electron charge. Dispersion of Laponite primarily contains $Na^+$, $OH^-$, $H^+$ and $Cl^-$ ions, wherever applicable, that contribute to the ionic conductivity. However at the dispersion $pH_f$ of around 10.4, concentration of $H^+$ ions is very small to influence the conductivity. It is important to note that, leaching of $Mg^{2+}$ and $Li^+$ ions in 1 and 1.7 mass% dispersion have a very little effect on the ionic conductivity because of their small concentration compared to the other ions. Therefore, their effect is neglected in these calculations. Concentration of $OH^-$ ion can be directly obtained from the $pH_f$ of dispersion as $n_{OH} = 10^{pH_f - 14}$. The $Cl^-$ ions are present only for those dispersions that are prepared in water having acidic $pH_i$. Therefore concentration of $Cl^-$ ions is given by: $n_{Cl} = 10^{-pH_i}$ (for $pH_i < 7$). The knowledge of mobilities of the mentioned ions ($\mu_{Na} = 5.19 \times 10^{-8}$ m$^2$/sV, $\mu_{Cl} = 7.908 \times 10^{-8}$ m$^2$/sV and $\mu_{OH} = 2.05 \times 10^{-7}$ m$^2$/sV) (Haynes, 2010) and conductivity ($\sigma$), therefore directly leads to estimation of concentration of $Na^+$ ions in the dispersion.

In figure 5, $n_{Na}$ has been plotted for the studied dispersions, wherein concentration of $Na^+$ ions can be seen to be increasing with increase in the concentration of dispersion, but is independent of $pH_i$ irrespective of whether it is acidic or basic if maintained either by HCl or NaOH. The concentration of $Na^+$ ions in 2.8 mass% dispersion is almost double than that of at 1.7 mass% dispersion, though at both the concentrations dispersions have similar $pH_f$. This result indicates that greater $n_{Na}$ might be having stabilizing effect on Laponite dispersion against $CO_2$ dissolution. In order to test this hypothesis further, two dispersions are prepared, which are 1 and 1.7 mass% Laponite in water having $pH_i = 7$ with 11.1 and 7.2 mM NaCl, respectively. The concentration of NaCl is decided in such a fashion that $n_{Na}$ in 1 and 1.7 mass% dispersion matches with that of 2.8 mass% dispersion. The corresponding $pH_f$ and $n_{Na}$ for these two dispersions having NaCl is plotted in figures 3 and 5 respectively, which can be seen to be closely matching with that of 2.8 mass% dispersion. Concentration of $Mg^{+2}$ ions is measured in both these dispersions (1 and 1.7 mass% having salt), and remarkably both the dispersions are observed to be free of any measurable amount of $Mg^{+2}$ ions. This experiment clearly indicates possible key role played by $Na^+$ ion concentration in stabilizing Laponite dispersion against dissolution. It should be noted that significant amount of $Na^+$ ions are also present in citric acid buffer. However, if $pH_f$ of Laponite dispersion is itself lesser than 9, the dissolution of Laponite cannot be avoided irrespective of the concentration of $Na^+$ ions. Therefore, in citric acid buffer dispersions high concentration of $Na^+$ ions cannot prevent dissolution of Laponite particles.

Interestingly the observations of the present work share some similarities with that of the experiments by Thompson and Butterworth (1992), who observe $pH_f$ of Laponite dispersion and $n_{Na}$ to increase simultaneously with increase in concentration of Laponite. They do not observe detectable concentration of $Mg^{+2}$ ions in dispersion having $n_{Na} > 0.7$ mM and $pH_f > 9$. However, owing to simultaneous increase in $pH_f$ and $n_{Na}$, it is difficult to separate the independent influence of the either in the work of Thompson and Butterworth (1992). In the present study, on the other hand, leaching of $Mg^{+2}$ ions is observed when concentration of $Na^+$ ions is low, but $pH_f$ is almost identical and above 10. Therefore, the proposal of greater concentration of $Na^+$ ions having a stabilizing effect on Laponite against dissolution is in qualitative agreement with that of Thompson and Butterworth (1992). However, Thompson and Butterworth (1992) also reported that concentration of NaCl has practically no effect on the dissolution of Laponite dispersion; in the present work, instead, it is observed that incorporation of NaCl in Laponite dispersion show a further stabilising effect. The reasons of discrepancy might be due to the differences in the procedures to manufacture Laponite dispersions used by Thompson and Butterworth (1992) and used in the present study. This is evident from that fact that Laponite used by Thompson and Butterworth (1992) has a diameter of 20 nm and thickness in the range 2 to 4 nm, while the present Laponite has diameter of 25 nm and thickness of 1 nm. In addition, there is also difference in procedure employed to estimate pH and concentration of $Mg^{+2}$ ions. In the present work the dispersion itself is tested for the pH measurement as well as for the titration to estimate the concentration of $Mg^{+2}$ ions. Thompson and Butterworth (1992), on the other hand, perform dialysis and carry out titration and pH measurement on dialysate solution. Overall, the present work clearly indicates that in order to prevent dissolution of Laponite particles in aqueous media, $pH_f$ of the dispersion is not the only criterion, but concentration of $Na^+$ ions also plays an important role.

## 4.    Conclusion

In this work stability of Laponite in an aqueous dispersion against dissolution is investigated which results in leaching of magnesium ions. For this, different aqueous dispersions are studied having different concentrations of Laponite (1, 1.7 and 2.8 mass%), different initial $pH_i$ of water before mixing Laponite (from pH 3 to 10), and different reagents to maintain initial $pH_i$ of water (citric acid buffer, HCl, and NaOH. When acidic $pH_i$ of water is maintained by citric acid buffer in the range 3 to 6, incorporation 2.8 mass% Laponite in the same is observed to raise $pH_f$ of the dispersions in between 7.5 and 9. For such dispersions, significant amount of magnesium leaching is observed. On the other hand, when HCl or NaOH is used to maintain $pH_i$ of water before mixing in the range 3 to 10, incorporation of Laponite in the same raises its $pH_f$ above 10 irrespective of the concentration of Laponite or the $pH_i$ of water. Such enhancement is attributed to dissociation of $OH^-$ ions from the edges of the Laponite particles.

However, very importantly, although the $pH_f$ of dispersion is above 10, leaching of magnesium ions is observed in low concentration dispersion (1 and 1.7 mass%) but not in high concentration dispersion (2.8 mass%). This observation is contrary to the usual belief that leaching of magnesium ions is possible only when $pH_f$ of dispersion is below 9. In order to investigate this intriguing behaviour, ionic conductivity of the dispersion is measured. Knowledge of the mobilities of various ions leads to concentration of $Na^+$ ions in the dispersion. It is observed that concentration of $Na^+$ ions in the dispersion increases with increase in concentration of Laponite but it is independent of $pH_i$ of water. Interestingly incorporation of salt (NaCl) in 1 and 1.7 mass% dispersions, wherein resultant concentration of $Na^+$ ions is same as that of 2.8 mass% dispersion, are also observed to prevent dissolution. Therefore, there is a possibility that greater concentration of $Na^+$ ions might have a role in rendering greater stability to the Laponite dispersions against dissolution.

**Acknowledgment:** The financial support from the department of Atomic Energy- Science Research Council, Government of India is greatly acknowledged.